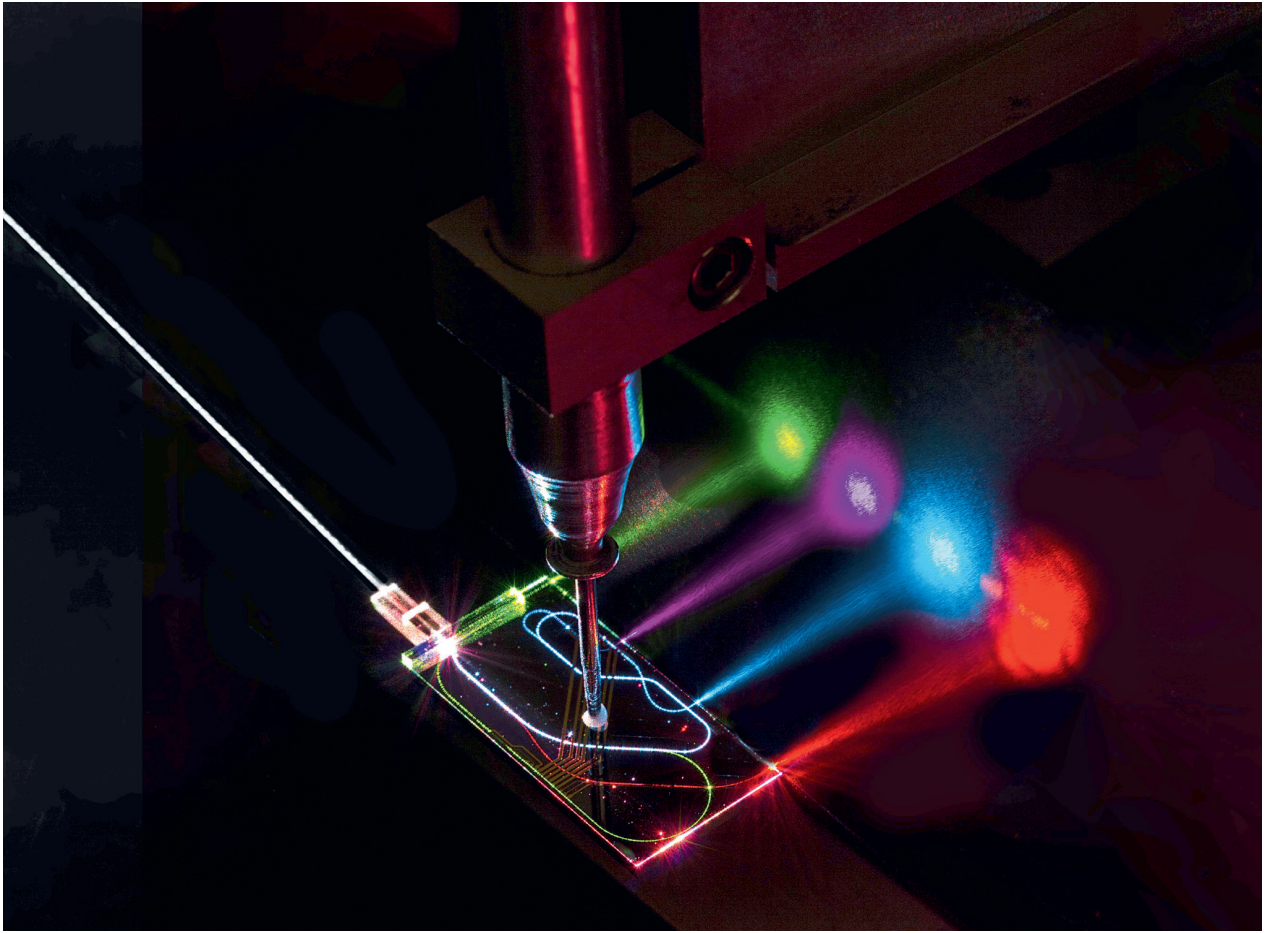

# Photonic integrated circuits for life sciences

A large number of discrete optical components could be replaced by a photonic integrated circuit in a multi-color laser engine for the visible spectral range. The photonic integrated circuit is based on silicon nitride waveguide technology.

Jeremy Witzens[1], Patrick Leisching[2], Alireza T. Mashayekh[1], Thomas Klos[2], Sina Koch[2], Florian Merget[1], Douwe Geuzebroek[3], Edwin Klein[3], Theo Veenstra[3], Ronald Dekker[3]
[1]RWTH Aachen University
[2]Toptica Photonics
[3]Lionix International

We report on the use of silicon nitride (SiN) photonic integrated circuits (PICs) in high-value instrumentation, namely multi-color laser engines (MLEs), a core element of cutting-edge biophotonic systems applied to confocal microscopy, fluorescent microscopy – including super-resolution stimulated emission depletion (STED) microscopy – flow cytometry, optogenetics, genetic analysis and DNA sequencing, to name just a few. These have in common the selective optical excitation of molecules – fluorophores, or, in the case of optogenetics, light-gated ion channels – with laser radiation falling within their absorption spectrum. Unambiguous identification of molecules or cellular subsets often requires jointly analyzing fluorescent signals from several fluorescent markers, so that MLEs are required to provide excitation wavelengths for several commercially available biocompatible fluorophores.

A number of functionalities are required from MLEs in addition to sourcing the required wavelengths [1]: variable attenuation and/or digital intensity modulation in the Hz to kHz range are required for a number of applications such as optical trapping, lifetime imaging or fluorescence recovery after photobleaching (FRAP). Moreover, switching of the laser between two fiber outputs can be utilized, for example, to switch between scanning confocal microscopy and wide-field illumination modes, for instance, for conventional fluorescence imaging.

MLEs assembled from discrete components require expensive components such as acousto-optic tunable filters (AOTFs) and fiber switches. Moreover, maintaining the long-term alignment of discretely assembled elements can be an engineering challenge and thus also a cost driver. In addition to reducing the bill of materials and the number of discrete elements that have to be assembled, miniaturization of MLEs by taking advantage of SiN PICs also removes constraints in the architecture of the entire system: rather than fiber coupling the MLE to the instrument, it would then be practicable to directly flange on the miniaturized MLE, minimizing the overall system size. This then opens the potential for further innovation, such as directly beam forming the light on the PIC prior to free-space coupling to the instrument.

Further miniaturization can be achieved by replacing diode-pumped solid-state (DPSS) lasers by semiconductor diode lasers wherever possible. While this also enables direct modulation of the laser, the need for external modulation/variable attenuation is not fully substituted, as turning down the diode lasers to very low power levels can also result in substantially increased noise. The capability of attenuating the laser radiation inside the PIC, even with a finite extinction, thus allows an increase in the minimum power levels to which lasers have to be driven and will thus result in an improved noise performance.

## PIC platform

Lionix International and Belgium's Interuniversity Microelectronics >

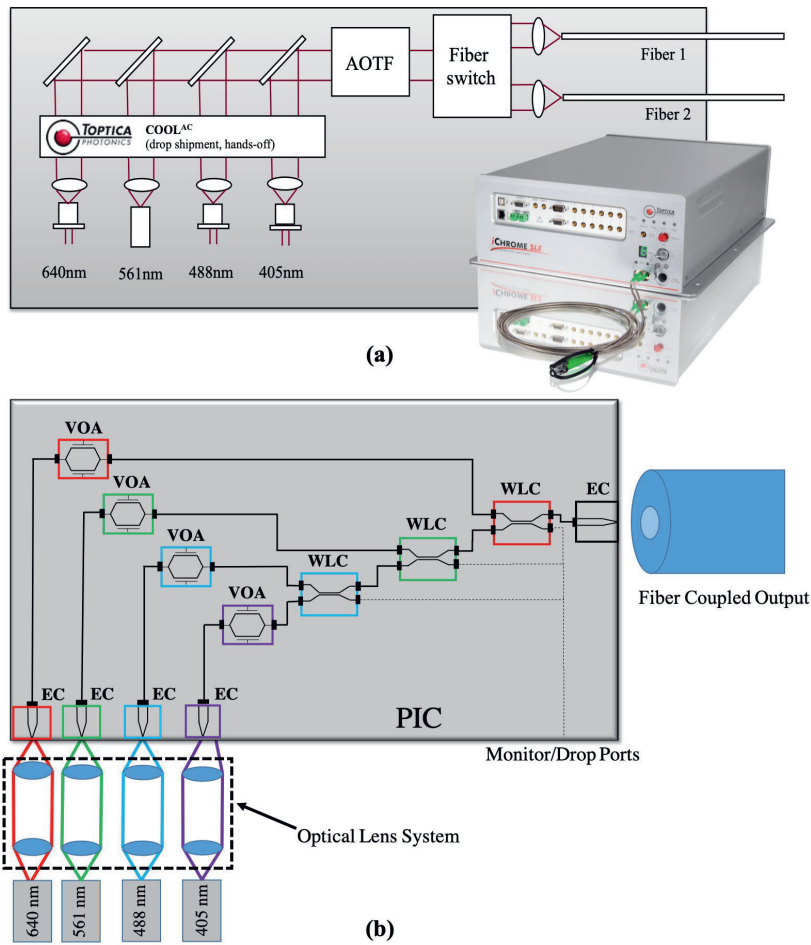

**Figure 1:** Schematics of (a) a commercially available MLE based on discrete components and (b) a SiN photonic integrated circuit (PIC) combining the key functionalities. On-chip devices consist of edge couplers (EC), variable optical attenuators (VOA) and wavelength combiners (WLC).

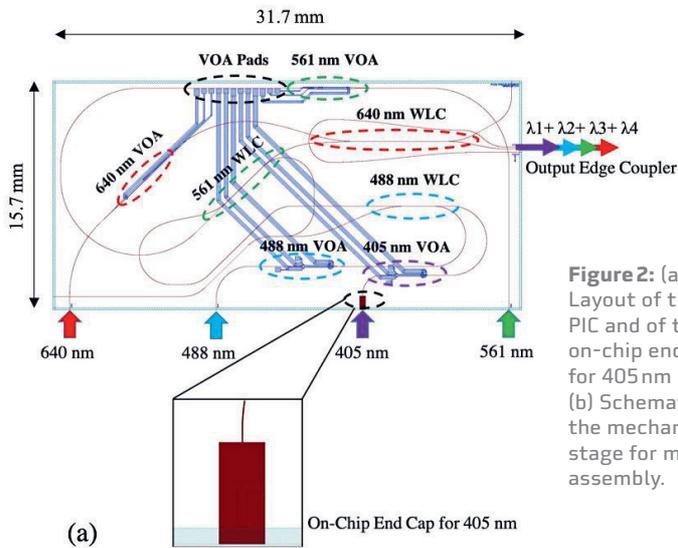

**Figure 2:** (a) Layout of the PIC and of the on-chip end cap for 405 nm (inset). (b) Schematic of the mechanical stage for module assembly.

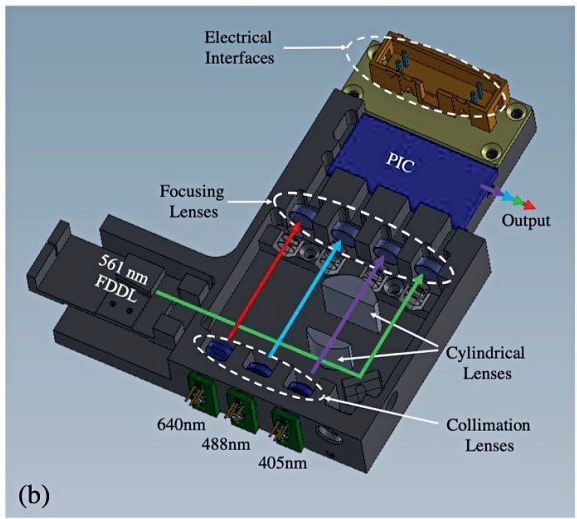

Centre (imec) have extended the capabilities of their SiN Triplex and Biopix platforms to the visible wavelength range in the framework of the project Pix4life, funded by the European Union to support the growing application space of SiN-based visible wavelength PICs in the Life Sciences. This platform is expected to be fully accessible through open access multi-project wafer (MPW) fabrication runs in 2020, providing a convenient and cost-effective access to this PIC technology.

The work reported in this article has been implemented in the Triplex platform. The currently available visible wavelength platform consists of relatively low-confinement waveguides formed by fully etching a single stoichiometric $Si_3N_4$ film deposited by low-pressure vapor deposition (LPCVD) onto oxidized silicon wafers and later top-clad with deposited $SiO_2$ (the single stripe geometry). The material system has a wide transparency range starting at 400 nm and reaching up to 2.35 µm, as limited by absorption through $SiO_2$ for the larger wavelengths. The lower wavelength cut-off is very sensitive to material quality, as non-stoichiometric silicon rich materials in particular result in higher absorption at the shorter wavelengths. Thus, this platform enables coverage of the entire visible range used to excite fluorophores commonly utilized in the life sciences.

As a particularity of the technology, SiN films are relatively thin (few tens of nm), resulting in reduced confinement so that relatively large bending radii are re-

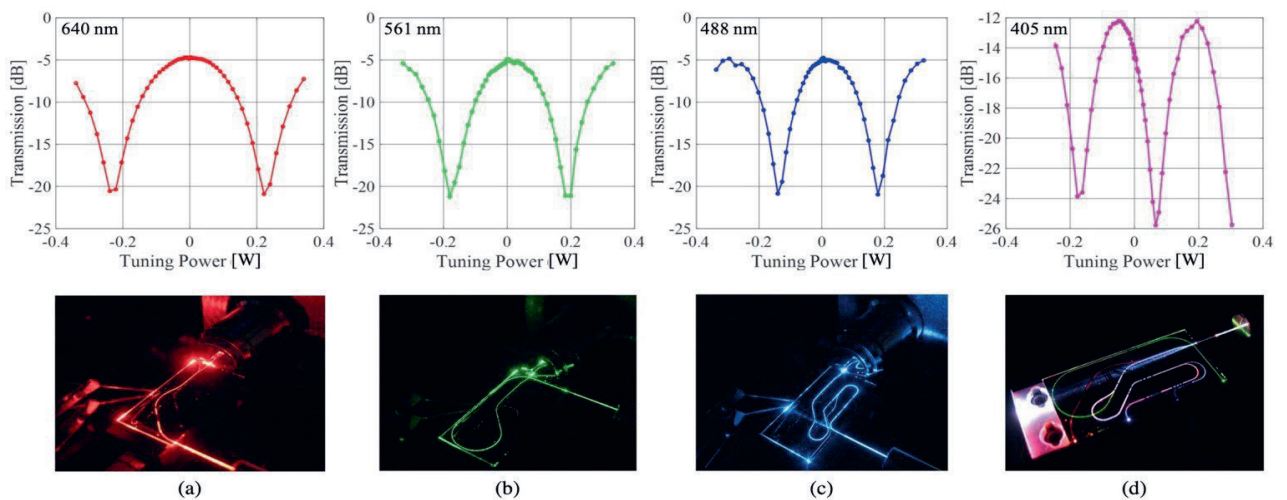

**Figure 3:** Top: Transmission through the PIC (one fiber-coupled interface) for (a) 640 nm, (b) 561 nm, (c) 488 nm, and (d) 405 nm. Bottom: Photograph of the PIC with the corresponding wavelengths injected. In (d) all the wavelengths are switched on at the same time to visualize the different optical paths and the multiplexing functionality.

## Summary

While a significant body of work exists on the utilization of PIC technology for cheap consumables for lab-on-a-chip applications, here we focus on the utilization of visible wavelength PICs in high-value instrumentation. We have presented chip and system-level characterization results for a multi-color laser engine based on a PIC incorporating the required core functionalities. The chip was implemented in the Triplex silicon nitride technology in the framework of the PIX4life pilot line, with layer thicknesses optimized for visible wavelength life-science applications. The demonstrated chip combines variable optical attenuation and multiplexing of four wavelengths covering the entire visible spectrum (405, 488, 561, 640 nm) with moderate overall insertion losses (~5 dB chip level, ~6.5 dB system level) for three of the wavelengths (488, 561, 640 nm). Further work aims to improve both insertion losses and reliability of the 405 nm interfaces. Moreover, after a first demonstration of module assembly, the focus will now move to verifying and guaranteeing the long-term stability of the assembly. In order to leverage the full flexibility afforded by PIC integration, a version of the chip including beam shaping in its output stage for direct free-space coupling into a flow cytometer is also under development.

quired for low loss transmission, but also facilitating matching of edge couplers to the beams emitted or received by standard visible wavelength optical fibers. While, without further material integration, the SiN/SiO$_2$ material system only allows relatively slow refractive index tuning via the thermo-optic effect [2], integration of ferroelectric materials such as lead zirconate titanate (PZT) and application of stress via piezoelectric properties [3] or direct refractive index change via the Pockels effect [4] has yielded faster modulation schemes. Here, thermal phase shifters adequate for the targeted kHz modulation speeds are utilized.

Details in regards to edge coupling to the developed PIC technology, a low-loss multiplexer device concept adjustable to target wavelengths in the entire visible range and capable of combining the targeted 405, 488, 561 and 640 nm wavelengths, as well as on-chip attenuators and switches can be found in the technical literature [2]. Here, we report on the performance characteristics of a complete MLE system chip combining multiplexing with wavelength-selective variable optical attenuation as well as post-assembly, module-level insertion losses.

A typical commercial MLE from Toptica is depicted in **figure 1(a)**. Four wavelengths – 405, 488, 561 and 640 nm – are combined with dichroic mirrors, modulated by an AOTF and sent through a fiber switch allowing rapid switching between different illumination modes. Constant optical output levels (COOL) with auto-calibration (AC) are achieved with Toptica's proprietary COOL$^{AC}$ units that allow adaptive steering of the beam. **Figure 1(b)** is a representation of the targeted MLE implementation with a SiN PIC. Light is coupled through focusing lenses to input edge couplers. On-chip, thermally-tuned variable optical attenuators (VOAs) adjust the light intensity independently for each wavelength, after which the four are multiplexed by means of cascaded wavelength combiners (WLC). In an extended version of the chip, the VOAs have two complementary output ports that are sent to independent sets of wavelength combiners and coupled to one out of two fibers at the output, fulfilling all the functionalities of the commercial MLE represented in **figure 1(a)**. While the lasers are free space coupled to the input of the chip with a pair of lenses, light is directly edge coupled to fibers at the output by attaching a fiber array to the optical subassembly [5].

## PIC layout and module concept

**Figures 2(a)** and **2(b)** show the layout of the chip introduced above as well as the mechanical design of the full module. Beams from the four lasers are individually free-space coupled to the PIC with help of a lens system. Three wavelengths (405, 488 and 640 nm) are directly generated by semiconductor laser diodes. The remaining one, 561 nm, is generated by a 1122 nm laser diode whose emission is subsequently frequency doubled (referred to as a fre- >

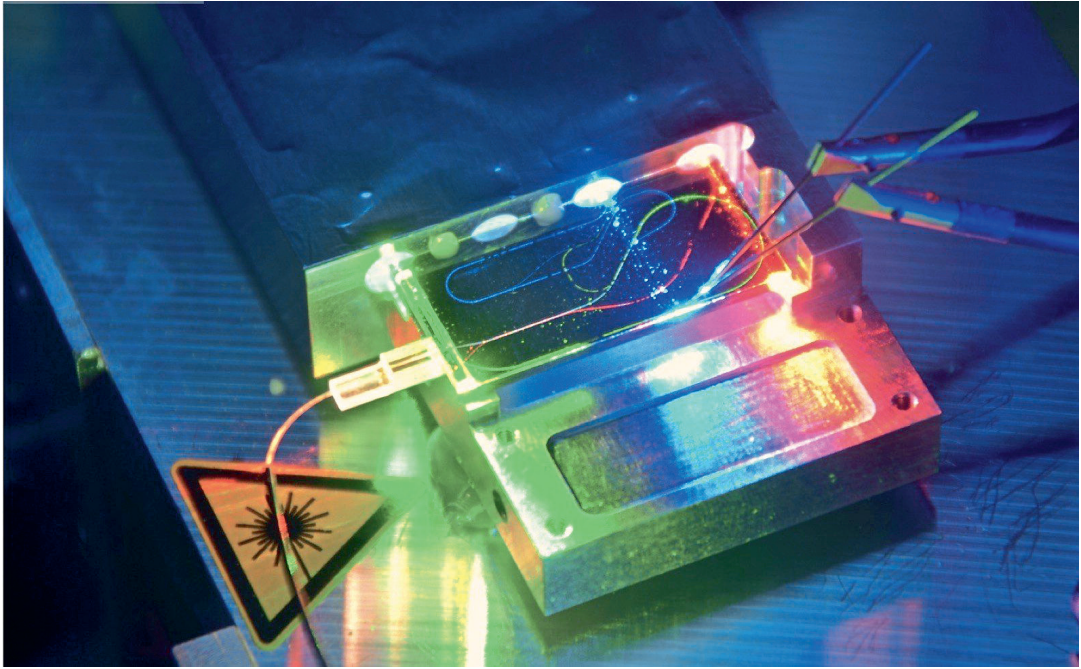

**Figure 4:** Photograph of the assembled module under operation. Individual wavelengths are coupled into the PIC at input ports located along the long side of the chip towards the top of the picture. After multiplexing, they are all coupled to a single fiber encased in a fiber array attached to the short side of the chip, on the left side.

quency doubled diode laser, FDDL). An additional mirror serves to steer the beam from the FDDL to the PIC. For 488, 561 and 640 nm, a pair of convex lenses prove sufficient to collimate the beam and later focus it onto the corresponding edge coupler.

For 405 nm an endcap interface is under development. Chip and fiber interfaces exposed to near-UV short-wavelength high-intensity radiation suffer from rapid degradation due to photo-assisted contamination, in particular via carbon contamination. To slow down facet degradation, optical fibers can be fused to a short section of coreless fiber – the endcap. The beam can then be focused to the point inside the fiber at which the coreless section transitions to the single-mode core. Optical intensities at the air interface can be reduced two to three orders of magnitude, greatly improving the reliability of the fiber. A similar concept will be applied to the 405 nm interface in order to improve PIC reliability.

The order of the input interfaces was constrained by the requirement to route the longer wavelengths 561 and 640 nm with larger waveguide bends to avoid bending losses, the corresponding routing being facilitated by allocating them to the outer input ports (see **figure 2(a)**). The wavelengths 561 and 640 nm were routed with minimum bending radii of respectively 5.4 and 7 mm. The shorter wavelengths 405 and 488 nm were routed with minimum bending radii of 1.3 and 1.7 mm. Since the resulting waveguide routing when using the low-loss optimized, low-confinement waveguides is one of the constraints in regards to the size of the chip, preventing further PIC miniaturization, Lionix has developed a high-confinement waveguide that allows tighter bends and is planning to additionally offer this within the MPW framework.

### System chip measurements

The PIC was characterized using Toptica's iChrome MLE sourcing the four required wavelengths. After rotating the polarization to the TM polarization required by the PIC design, light was directly end-fire coupled from a cleaved single-mode fiber to one of the four input interfaces of the chip (while input edge couplers were meant for free space coupling, they have almost the same waveguide width as the output edge coupler, so that resulting insertion losses are identical to the regular packaging scheme with a fiber at the output). The utilized fiber, a Panda-type polarization maintaining fiber with a core diameter adjusted to maintain single-mode operation down to 400 nm (Nufern part number PM-S405-XP), has a $1/e^2$ mode-field diameter (MFD) ranging from 3.3 µm at 405 nm to 4.6 µm at 630 nm. At the output of the PIC, light is collected by a microscope objective, transmitted through a polarization beam splitter to monitor and maintain polarization, and finally monitored by a wide area detector.

Insertion losses of the receiving apparatus were evaluated by measuring directly the power of a collimated laser beam without interposed PIC and subsequently normalized out from the raw PIC transmission data. As a consequence of the output light collection apparatus not being single mode, insertion losses due to mode mismatch are only incurred once (at the input interface), as opposed to the module in which mode mismatch will play a role at both the input and the output of the PIC. Measurements for 488, 561 and 640 nm were made by injecting light in the corresponding nominal input port and collecting the light from the PIC output. Since the end cap for the 405 nm input port was still under development at the time at which these measurements were taken, 405 nm transmission experiments were done in reverse direction, injecting light into the nominal output port and collecting it from the nominal input. During measurements, the voltage applied to the VOAs was swept from -15 V to +15 V, in order to identify the maximum transmission point.

Results are shown in **figure 3**. Only one phase shifter is actuated in the MZI-based VOAs at a given time. On the x-axis, a negative power represents a positive power applied to one of the phase shifters, while a positive power represents a power applied to the opposite phase shifter. For the three wavelengths 488, 561 and 640 nm, the overall PIC transmission losses are close to 5 dB, while measured losses at 405 nm are larger at 12 dB. These losses correspond to the cumulative effect of the input coupling losses with the insertion losses of the VOA and the cascaded WLCs and are well in line with device-level characterization results [2]. One of the difficulties in this PIC design is that a single output port serves to outcouple all four wavelengths, as they are being multiplexed on the chip. Consequently, the edge coupler geometry is not optimum for all the wavelengths and results from a trade-off. Here, it was optimized for the longer wavelengths 488, 561 and 640 nm, resulting in increased insertion losses at 405 nm. Retargeting the edge coupler width would allow decreasing the losses at 405 nm by 4 dB at the cost of an additional 1-2 dB at the other three wavelengths, resulting in more balanced losses [2]. **Figure 3** also shows photographs of the PIC visualizing the waveguides during forward injection of 640 nm (red), 561 nm (green) or 488 nm (blue) light. As previously explained, measurements at 405 nm (purple) could only be made in the reverse direction. **Figure 3(d)** shows a photograph of the chip with all four wavelengths injected through the nominal output port, illustrating the (de-)multiplexing functionality.

### Module characterization

A photograph of the assembled module under operation can be seen in **figure 4**. The overall insertion losses from the diode lasers to the fiber were determined to be -5, -7.8 and -6.6 dB, respectively for 488, 561 and 640 nm. These are on average about 1.5 dB higher than the transmission measured from the bare chip, and are well in line with expectations. Indeed, the chip was measured with single-mode fiber coupling on one interface, but collection by a large area detector on the other – with the latter adding virtually no losses. Here, on the other hand, light is in-coupled to the single-mode waveguides with a pair of collimating and focusing lenses, while also out-coupled into single-mode fibers. The modeled losses for the first interface are well in line with these additional 1.5 dB.

### Acknowledgements

The authors would like to acknowledge funding through the PIX4life project, funded by the EU under the H2020 program (contract number 688519).   mg ■

### Web service

Further details on the iChrome MLE series from Toptica

**www.photonik.de/33392**

### Contact

Prof. Dr. Jeremy Witzens
Chair of Integrated Photonics
RWTH Aachen University
Sommerfeldstr. 18/24
52074 Aachen, Germany
Tel. +49 241 80 20020
Fax +49 241 80 22212
jwitzens@iph.rwth-aachen.de
www.iph.rwth-aachen.de

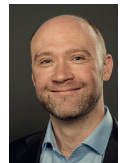

Dr. Patrick Leisching
Vice President
Research & Development
TOPTICA Photonics AG
Lochhamer Schlag 19
82166 Graefelfing
Germany
Tel. +49 89 85837-0
Fax +49 89 85837-200
patrick.leisching@toptica.com
www.toptica.com

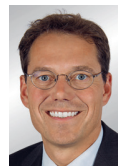